\documentclass[conference]{IEEEtran}
\IEEEoverridecommandlockouts
\usepackage{cite}
\usepackage{amsmath,amssymb,amsfonts}
\usepackage[dvips]{graphics}
\usepackage{graphicx}
\usepackage{textcomp}
\usepackage{hyperref}

\usepackage{xcolor}
\def\BibTeX{{\rm B\kern-.05em{\sc i\kern-.025em b}\kern-.08em
    T\kern-.1667em\lower.7ex\hbox{E}\kern-.125emX}}
\usepackage{setspace}

\usepackage{algorithm}

\usepackage[noend]{algpseudocode}
\newcommand{\algrule}[1][.2pt]{\par\vskip.5\baselineskip\hrule height #1\par\vskip.5\baselineskip}

{}

\usepackage{psfrag}

\begin{document}
\title{Capacity of Continuous Channels with Memory\\via Directed Information Neural Estimator
}

\author{\IEEEauthorblockN{Ziv Aharoni}
\IEEEauthorblockA{{Ben Gurion University} \\
zivah@post.bgu.ac.il}
\and
\IEEEauthorblockN{Dor Tsur }
\IEEEauthorblockA{{Ben Gurion University} \\
dortz@post.bgu.ac.il}
\and
\IEEEauthorblockN{Ziv Goldfeld}
\IEEEauthorblockA{{Cornell University} \\
goldfeld@cornell.edu}
\and
\IEEEauthorblockN{Haim H. Permuter}
\IEEEauthorblockA{{Ben Gurion University} \\
haimp@bgu.ac.il}
}

\maketitle

\begin{abstract}
Calculating the capacity (with or without feedback) of channels with memory and continuous alphabets is a challenging task. It requires optimizing the directed information (DI) rate over all channel input distributions. The objective is a multi-letter expression, whose analytic solution is only known for a few specific cases. When no analytic solution is present or the channel model is unknown, there is no unified framework for calculating or even approximating capacity. This work proposes a novel capacity estimation algorithm that treats the channel as a `black-box', both when feedback is or is not present. The algorithm has two main ingredients: (i) a neural distribution transformer (NDT) model that shapes a noise variable into the channel input distribution, which we are able to sample, and (ii) the DI neural estimator (DINE) that estimates the communication rate of the current NDT model. These models are trained by an alternating maximization procedure to both estimate the channel capacity and  obtain an NDT for the optimal input distribution. The method is demonstrated on the moving average additive Gaussian noise channel, where it is shown that both the capacity and feedback capacity are estimated without knowledge of the channel transition kernel. The proposed estimation framework opens the door to a myriad of capacity approximation results for continuous alphabet channels that were inaccessible until now.
\end{abstract}

\section{Introduction}

Many discrete-time continuous-alphabet communication channels involve correlated noise or inter-symbol interference (ISI). Two predominant communication scenarios over such channels are when feedback from the receiver back to the transmitter is or is not present. The fundamental rates of reliable communication over such channels are, respectively, the feedback (FB) and feedforward (FF) capacity. Starting from the latter, the FF capacity of an $n$-fold point-to-point channel $P_{Y^n|X^n}$, denoted  $C_{\mathsf{FF}}$, is given by \cite{BOOK:Gallager}
\begin{equation}
    C_{\mathsf{FF}}= \lim_{n \rightarrow \infty}  \sup_{P_{X^n}}{\frac{1}{n} I(X^n;Y^n)}.\label{EQ:CFF}
\end{equation}
In the presence of feedback, the FB capacity $C_\mathsf{FB}$ is \cite{PAPER:CFB_formula_Kim}
\begin{equation}
    C_{\mathsf{FB}} = \lim_{n \rightarrow \infty}  \sup_{P_{X^n \| Y^{n-1}}}{\frac{1}{n} I(X^n \rightarrow Y^n)}\label{EQ:CFB}
\end{equation}
where,
\begin{equation}\label{EQ:DI_sum}
    I(X^n \rightarrow Y^n) := \sum_{i=1}^n{I(X^i;Y_i|Y^{i-1})}   
\end{equation}
is the directed information (DI) from the input sequence $X^n$ to the output $Y^n$ \cite{PAPER:DI_Massey}, and $P_{X^n \| Y^{n-1}} := \prod_{i=1}^n P_{X_i|X^{i-1}Y^{i-1}}$ is the distribution of $X^n$ causally-conditioned on $Y^{n-1}$ (see \cite{PAPER:DI_Haim, THESIS:Kramer} for further details). Built on \eqref{EQ:DI_sum}, for stationary processes, the DI rate is defined as
\begin{equation}\label{DI_rate}
    I(\mathcal{X}\rightarrow\mathcal{Y}) := \lim_{n\rightarrow \infty}\frac{1}{n}I(X^n \rightarrow Y^n).
\end{equation}
As shown in \cite{PAPER:DI_Massey}, when feedback is not present, the optimization problem \eqref{EQ:CFB} (which amounts to optimizing over $P_{X^n}$ rather than $P_{X^n\|Y^n}$) coincides with \eqref{EQ:CFF}.
Thus, DI provides a unified framework for representing both FF and FB capacities. 

Computing $C_\mathsf{FF}$ and $C_\mathsf{FB}$ requires solving a multi-letter optimization problem. Closed form solutions to this challenging task are known only in several special cases. 
A common example for $C_\mathsf{FF}$ is the Gaussian channel with memory \cite{BOOK:Cover} and the ISI Gaussian channel \cite{PAPER:ISI_CFF_Massey}.
There are no known extensions of these solutions to the non-Gaussian case.
For $C_\mathsf{FB}$, a solution for the 1st order moving average additive Gaussian noise (MA(1)-AGN) channel was found \cite{PAPER:MA_Kim}. Another closed form characterization is available for auto-regressive moving-average (ARMA) AGN channels \cite{PAPER:ARMA_Yang}. To the best of our knowledge, these are the only two non-trivial examples of continuous channels with memory whose FB capacity is known in closed form. Furthermore, when the channel model is unknown, there is no numerically tractable method for approximating capacity based on samples. 

Recent progress related to capacity computation via deep learning (DL) was made in \cite{PAPER:DL_MINE}, where the mutual information neural estimator (MINE) \cite{PAPER:MINE} was used to learn modulations for memoryless channels. Later, \cite{PAPER:ZIV_ISING} proposed an estimator based on a reinforcement learning algorithm that iteratively estimates and maximizes the DI rate was proposed, but only for discrete alphabet channels with a known channel model.

Inspired by the above, we develop the framework for estimating FF and FB capacity of arbitrary continuous-alphabet channels, possible with memory, without knowing the channel model.
Our method does not need to know the channel transition kernel.
We only assume a stationary channel model and that channel outputs can be sampled by feeding it with inputs.
Central to our method are a new DI neural estimator (DINE), used to evaluate the communication rate,
and a neural distribution transformer (NDT), used to simulate input distributions. Together, DINE and NDT lay the groundwork for our capacity estimation algorithm. In the remainder of this section, we describe DINE, NDT, and their integration into the capacity estimator.\\

\subsection{Directed Information Neural Estimation}

The estimation of mutual information (MI) from samples using neural networks (NNs) is a recently proposed approach \cite{PAPER:MINE, PAPER:MINEE}. It is especially effective when the involved random variables (RVs) are continuous.
The concept originated from \cite{PAPER:MINE}, where MINE was proposed. The core idea is to represent MI using the Donsker-Varadhan (DV) variational formula
\begin{equation}\label{EQ:DV_MINE}
    I(X;Y) = \sup_{\mathsf{T}: \mathcal{X} \times \mathcal{Y} \rightarrow \mathbb{R}}{\mathbb{E}\left[ \mathsf{T}(X,Y) \right]} - \log\mathbb{E}\left[ e^{\mathsf{T}(\widetilde{X},\widetilde{Y})} \right],
\end{equation}
where $(X,Y) \sim P_{XY}$ and $(\widetilde{X},\widetilde{Y}) \sim P_X \otimes P_Y$. The supremum is over all measurable functions $\mathsf{T}$ for which both expectations are finite. Parameterizing  $\mathsf{T}$ by an NN and replacing expectations with empirical averages, enables gradient ascent optimization to estimate $I(X;Y)$. A variant of MINE that goes through estimating the underlying entropy terms was proposed in \cite{PAPER:MINEE}. The new estimators were shown empirically to perform extremely well, especially for continuous alphabets.



Herein, we propose a new estimator for the DI rate $I(\mathcal{X} \rightarrow \mathcal{Y})$.
The DI is factorized as 
\begin{equation}\label{EQ:DI_entropy}
    I(X^n \rightarrow Y^n) = h(Y^n) - h(Y^n\| X^n),
\end{equation}
where $h(Y^n)$ is the differential entropy of $Y^n$ and $h(Y^n\| X^n) := \sum_{i=1}^n{h(Y_i|Y^{i-1}, X^i)}$. Applying the approach of \cite{PAPER:MINEE} to the entropy terms, we expand each as a Kullback-Leibler (KL) divergence plus a cross-entropy (CE) residual and invoke the DV representation. To account for memory, we derive a formula valid for causally dependent data, which involves RNNs as function approximators (rather than the FF network used in the independently and identically distributed (i.i.d.) case). Thus, DINE is an RNN-based estimator for the DI rate from $X^n$ to $Y^n$ based on their samples.

Estimation of DI between discrete-valued processes was studied in \cite{PAPER:Haim_DI_est, PAPER:Other_DI_est, PAPER:DI_Neural_Spikes}. An estimator of the transfer entropy, which upper bounds DI for jointly Markov process with finite memory, was proposed \cite{PAPER:TE}. DINE, on the other hand, does not assume Markovity nor discrete alphabets, and can be applied to continuous-valued stationary and ergodic processes. A detailed description of the DINE algorithm is given in subsection \ref{DINE_sub_method}.


\subsection{Neural Distribution Transformer and Capacity Estimation}

\begin{figure}[ht!]{
    \psfrag{A}[][][0.65]{$N_i$} \psfrag{B}[][][0.8]{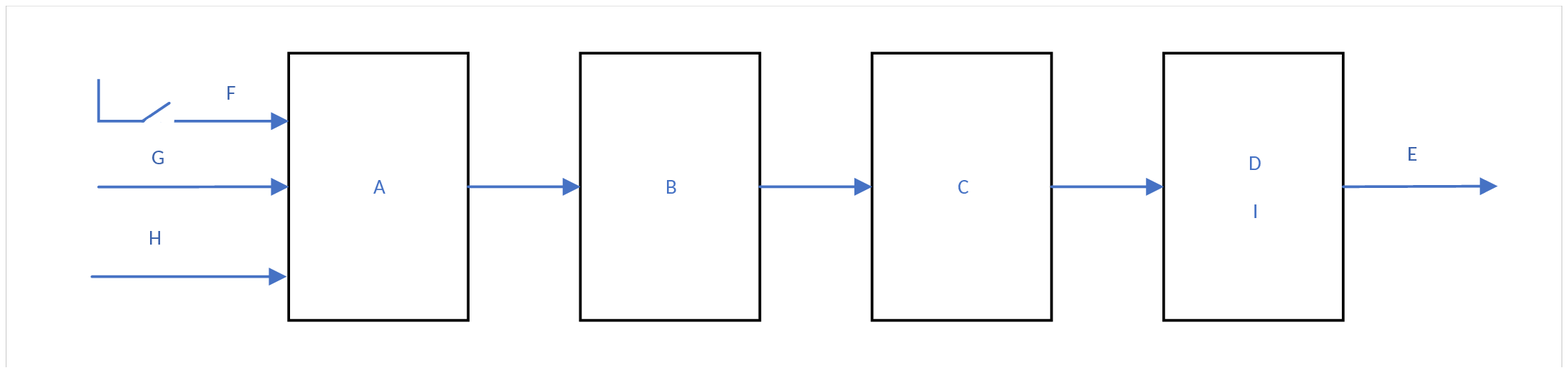} \psfrag{C}[][][0.65]{Channel} \psfrag{D}[][][0.8]{DINE} \psfrag{E}[][][0.8]{$X_i$} \psfrag{G}[][][0.7]{$\widehat{I}_{\mathcal{D}_n}(X^n \rightarrow Y^n)$} \psfrag{H}[][][0.8]{$Y_i$} \psfrag{I}[][][0.8]{(RNN)} \psfrag{J}[][][0.8]{Feedback} \psfrag{K}[][][0.8]{(RNN)} \psfrag{L}[][][0.75]{$P_{Y_i|X^i Y^{i-1}}$} \psfrag{Z}[][][0.75]{Noise} \psfrag{M}[][][0.75]{$\Delta$} \psfrag{N}[][][0.75]{Output} \psfrag{O}[][][0.65]{$\widehat{I}_{\mathcal{D}_n}(\mathcal{X} \rightarrow \mathcal{Y})$} \psfrag{P}[][][0.68]{Gradient} \psfrag{Q}[][][0.75]{$Y_{i-1}$}
    \psfrag{R}[][][0.75]{(RNN)}
    \centerline{\includegraphics[scale=0.425]{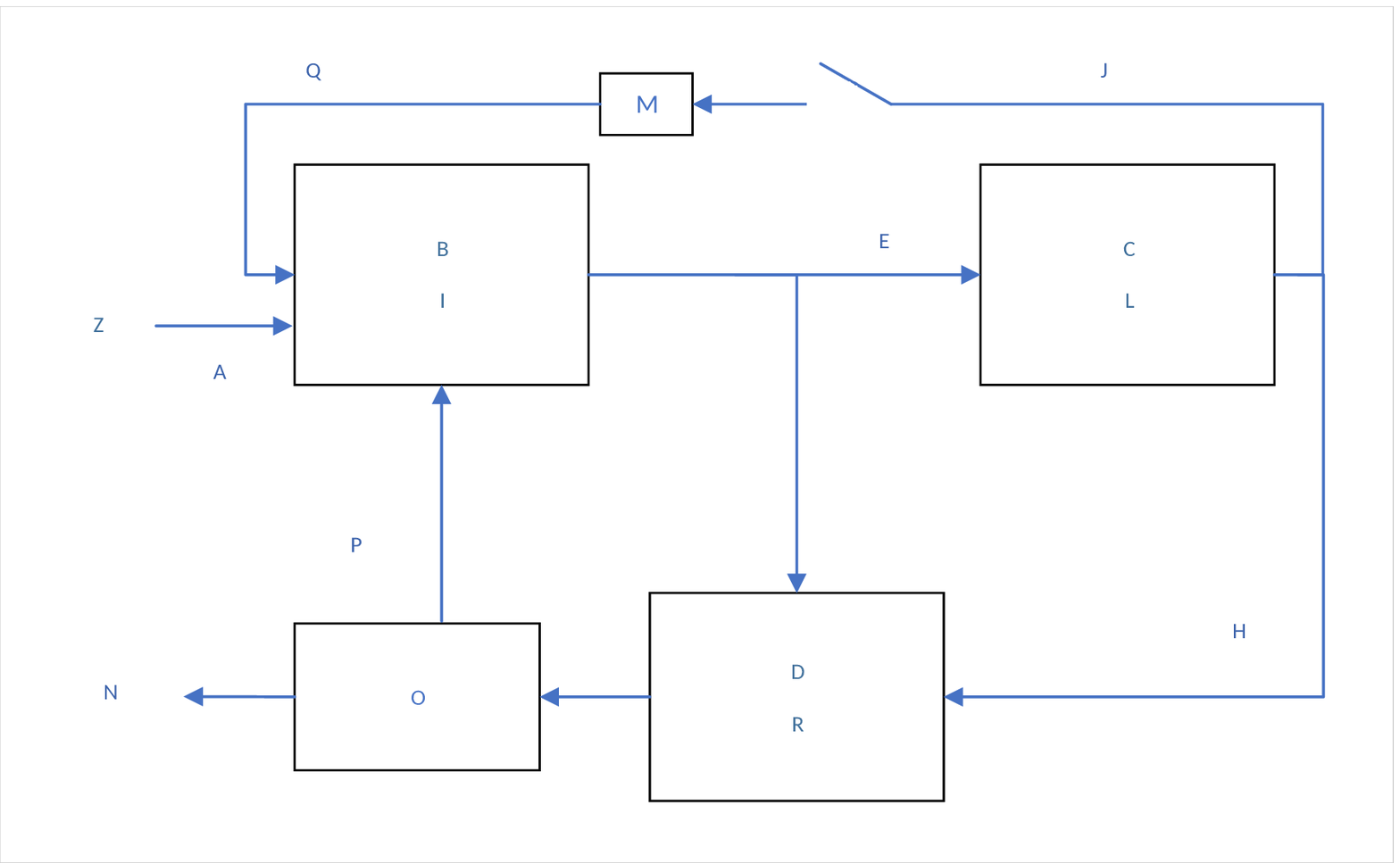}}}
    \caption{The overall capacity estimator: NDT generates samples that are fed into the channel. DINE uses these samples to improve its estimation of the communication rate. DINE then supplies gradient for the optimization of NDT.}
    \label{fig:complete_system}
\end{figure}

DINE accounts for one of the two tasks involved in estimating capacity, it estimates the objective of \eqref{EQ:CFB}. It then remains to optimize this objective over input distributions. To that end, we design a deep generative model, termed the NDT, to approximate the channel input distributions. This is similar in flavor to generators used in generative adversarial networks \cite{PAPER:GAN}.
The designed NDT maps i.i.d. noise into samples of the channel input distribution. For estimating FB capacity, in addition to the i.i.d. noise, the NDT also receives channel FB as inputs.
Together, NDT and DINE form the overall system that estimates the capacity as shown in Fig~\ref{fig:complete_system}.

The capacity estimation algorithm trains DINE and NDT models together via an alternating optimization procedure (i.e., fixing the parameters of one model while training the other). DINE estimates the communication rate of a fixed NDT input distribution, and the NDT is trained to increase its rate with respect to fixed DINE model. Proceeding until convergence, this results in the capacity estimate, as well as an NDT generative model for the achieving input distribution. We demonstrate our method on the MA(1)-AGN channel. Both $C_{\mathsf{FF}}$ and $C_{\mathsf{FB}}$ are estimated using the same algorithm, using the channel as a black-box to solely generate samples. The estimation results are compared with the analytic solution to show the effectiveness of the proposed approach.

\section{Methodology}
We give a high-level description of the algorithm and its building blocks. Due to space limitations, full details are reserved to the extended version of this paper. The implementation is available on GitHub.\footnote[2]{\url{https://github.com/zivaharoni/capacity-estimator-via-dine}}
\subsection{Directed Information Estimation Method}\label{DINE_sub_method}
We propose a new estimator of the DI rate between two correlated stationary processes, termed DINE.
Building on \cite{PAPER:MINEE}, we factorize each term in  \eqref{EQ:DI_entropy} as:

 \begin{align}\label{EQ:Entropy_decomposition}
    h(Y^n) &= h_{\mathsf{CE}}(P_{Y^n},P_{Y^{n-1}}\otimes P_{\widetilde{Y}}) \nonumber\\
    &\quad\quad\quad\quad\quad\quad\quad\ \ - D_{\mathsf{KL}}(P_{Y^n}\| P_{Y^{n-1}}\otimes P_{\widetilde{Y}})\nonumber\\
    h(Y^n \| X^n) &= h_{\mathsf{CE}}\left(P_{Y^n \| X^n},P_{Y^{n-1} \| X^{n-1}}\otimes P_{\widetilde{Y}}\middle |P_{X^n}\right) \nonumber\\
    &\quad- D_{\mathsf{KL}}\left(P_{Y^n\| X^n}\middle\| P_{Y^{n-1}\| X^{n-1}}\otimes P_{\widetilde{Y}}\middle |P_{X^n}\right)
\end{align}
where $h_{\mathsf{CE}}(P_X,Q_X)$ and $D_{\mathsf{KL}}(P_X \| Q_X)$ are, respectively, the CE and KL divergence between $P_X$ and $Q_X$, with 
\begin{align}
    h_{\mathsf{CE}}(P_{Y|X},Q_{Y|X}|P_{X})&\mspace{-2mu}:=\int_{\mathcal{X}}\mspace{-9mu}h_{\mathsf{CE}}(P_{Y|X=x},Q_{Y|X=x})\mathrm{d} P_{X}(x)\nonumber\\
    D_{\mathsf{KL}}(P_{Y|X}\|Q_{Y|X}|P_{X})&\mspace{-2mu}:=\int_{\mathcal{X}}\mspace{-9mu}D_{\mathsf{KL}}(P_{Y|X=x}\|Q_{Y|X=x})\mathrm{d} P_{X}(x)
\end{align}
denoting their conditional versions; and $P_{\widetilde{Y}}$ is uniform reference measure over the support of the dataset.
To simplify notation, we use the shorthands
\begin{align}
    D_Y^{(n)} &:= D_{\mathsf{KL}}(P_{Y^n}\| P_{Y^{n-1}}\otimes P_{\widetilde{Y}}) \nonumber\\
    D_{Y \| X}^{(n)} &:=D_{\mathsf{KL}}(P_{Y^n\| X^n}\| P_{Y^{n-1}\| X^{n-1}}\otimes P_{\widetilde{Y}}).
\end{align}
Subtracting both elements in \eqref{EQ:Entropy_decomposition} and observing that the difference of 
CE terms equals the DI at the former time step, we have
\begin{equation}\label{EQ:recursive_DI}
    I(X^n \rightarrow Y^n) = I(X^{n-1}\rightarrow Y^{n-1}) + D_{Y \| X}^{(n)} - D_Y^{(n)}. 
\end{equation}
Note that the difference of KL divergences equals $I(X^n;Y_n|Y^{n-1})$. For stationary data processes we take the limit and obtain
\begin{equation}
    \lim_{n \rightarrow \infty}D_{Y \| X}^{(n)} - D_Y^{(n)} = \lim_{n \rightarrow \infty}I(X^n;Y_n|Y^{n-1}) = I(\mathcal{X}\rightarrow \mathcal{Y}).
\end{equation}

Each $D_{\mathsf{KL}}$ is expanded by its DV representation \cite{PAPER:DV} as:

\begin{align}
    D_Y^{(n)} &= \sup_{\mathsf{T}:  \Omega \rightarrow \mathbb{R}}{\mathbb{E}\left[ \mathsf{T}(Y^n) \right]}
      - \log\mathbb{E}\left[ e^{\mathsf{T}(Y^{n-1},\widetilde{Y})} \right]\nonumber\\
      D_{Y\|X}^{(n)} &= \sup_{\mathsf{T}:  \Omega \rightarrow \mathbb{R}}{\mathbb{E}\left[ \mathsf{T}(Y^n \| X^n) \right]}
      - \log\mathbb{E}\left[ e^{\mathsf{T}(Y^{n-1}\|X^{n-1},\widetilde{Y})} \right]\label{EQ:DV_DINE}.
\end{align}
To maximize \eqref{EQ:DV_DINE}, each DV potential is parametrized by a modified LSTM and expected values are estimated by empirical averages over the dataset $\mathcal{D}_n := \{ (x_i,y_i) \}_{i=1}^n$.
Thus, the optimization objectives are:
\begin{align}\label{EQ:estimator_explicit}
    \widehat{D}_{Y \| X}(\theta_{Y\|X}, \mathcal{D}_n) &:= \frac{1}{n}\sum_{i=1}^{n}{\mathsf{T}_{\theta_{Y\| X}}(y_i|x^{i}y^{i-1})} \nonumber\\
        &- \log\left(\frac{1}{n}\sum_{i=1}^{n}{e^{\mathsf{T}_{\theta_{Y\| X}}(\widetilde{y_i}|x^{i}y^{i-1})}} \right) \nonumber\\
    \widehat{D}_{Y}(\theta_{Y}, \mathcal{D}_n) &:= \frac{1}{n}\sum_{i=1}^{n}{\mathsf{T}_{\theta_{Y}}(y_i|y^{i-1})} \nonumber\\
        &- \log\left(\frac{1}{n}\sum_{i=1}^{n}{e^{\mathsf{T}_{\theta_{Y}}(\widetilde{y_i}|y^{i-1})}} \right)
\end{align}
where $\tilde{y}^n \overset{\text{i.i.d.}}{\sim} P_{\widetilde{Y}}$ and $ \mathsf{T}_{\theta_{Y}}$, $ \mathsf{T}_{\theta_{Y\| X}}$ are the parametrized potentials.

The estimator is given by:
\begin{equation}\label{EQ:DI_est}
    \widehat{I}_{\mathcal{D}_n}(\mathcal{X} \rightarrow \mathcal{Y}):= \sup_{\theta_{Y \| X} \in \Theta_{Y \| X}}{ \widehat{D}_{Y\| X}} - \sup_{\theta_{Y} \in \Theta_{Y}}{ \widehat{D}_{Y}}
\end{equation}
 By universal approximation of RNNs \cite{PAPER:Uni_App_LSTM} and Breiman's theorem \cite{PAPER:Breiman}, the maximizer of \eqref{EQ:DI_est} approaches $I(\mathcal{X} \rightarrow \mathcal{Y})$ as the number of samples grows, provided the neural networks are sufficiently expressive.

\begin{algorithm}[ht]
\caption{Directed Information Rate Estimation}
\label{DINE_algorithm}
\textbf{input:} Samples of the process $\mathcal{D}_n$.\\
\textbf{output:} $\widehat{I}_{\mathcal{D}_n}(\mathcal{X} \rightarrow \mathcal{Y})$, estimated directed information rate.
\algrule
\begin{algorithmic}
\State Initialize networks parameters $\theta_Y , \theta_{Y \| X}$.
\State \textbf{Step 1,} Optimization: 
\Repeat
\State  Draw a batch $\mathcal{D}_B = \{ (x_{(i-1)T}^{iT},y_{(i-1)T}^{iT}) \}_{i=1}^B$
\State  Feed the network with the examples and compute \\ \qquad loss $\widehat{D}_{Y \| X}(\theta_{Y\|X}, \mathcal{D}_B)$, $\widehat{D}_{Y}(\theta_{Y}, \mathcal{D}_B)$.

\State Update networks parameters:\\
\quad\quad$\theta_{Y\| X} \leftarrow \theta_{Y\| X} + \nabla\widehat{D}_{Y \| X}(\theta_{Y\|X}, \mathcal{D}_B)$\\
\quad\quad$\theta_{Y} \leftarrow \theta_{Y} +\nabla\widehat{D}_{Y}(\theta_{Y}, \mathcal{D}_B)$
\Until{convergence}

\State \textbf{Step 2,} Perfrom a Monte Carlo estimation over $\mathcal{D}_n$ and subtract loss evaluations to obtain estimation :
\hspace{\algorithmicindent}\hspace{\algorithmicindent}$\widehat{I}_{\mathcal{D}_n}(\mathcal{X} \rightarrow \mathcal{Y}) = \widehat{D}_{Y \| X}(\theta_{Y\|X}, \mathcal{D}_n) - \widehat{D}_{Y}(\theta_{Y}, \mathcal{D}_n)$ 
\end{algorithmic}
\end{algorithm}

To capture the time dependencies in $\mathcal{D}_n$ we introduce a modified LSTM network model for functional approximation.
LSTM \cite{PAPER:LSTM} is an RNN that receives a time series $\{y_i\}_{i=1}^T$ as input and for each $i$, performs a recursive non-linear transform to calculate its hidden state $s_i$. We denote the LSTM function by $F: (y_i,s_{i-1}) \longmapsto s_i$.
The full characterization of $F$ is provided in \cite{PAPER:LSTM}.

\begin{figure}[ht!]
    \psfrag{A}[][][0.8]{$\widetilde{S}_1$}\psfrag{B}[][][0.8]{$S_1$}\psfrag{C}[][][0.8]{$0$}\psfrag{D}[][][0.8]{$\widetilde{Y}_1$}\psfrag{E}[][][0.8]{$Y_1$}\psfrag{F}[][][0.8]{$F$}\psfrag{G}[][][0.8]{$F$}\psfrag{H}[][][0.8]{$\widetilde{S}_T$}\psfrag{I}[][][0.8]{$S_T$}\psfrag{J}[][][0.86]{$S_{T-1}$}\psfrag{K}[][][0.8]{$\widetilde{Y}_T$}\psfrag{L}[][][0.8]{$Y_T$}\psfrag{M}[][][0.8]{$F$}\psfrag{N}[][][0.8]{$F$}
    \centerline{\includegraphics[scale=0.3]{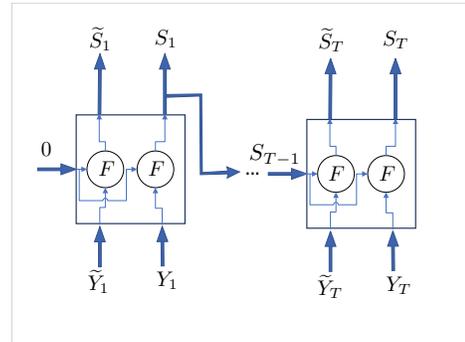}}
    
    \vspace{-4mm}
    \caption{The modified LSTM cell unrolled in the DINE architecture of $\widehat{D}_{Y}$. Recursively, at each time $i$, $(y_i, s_{i-1})$ and $(\widetilde{y}_i, s_{i-1})$ are mapped to $s_i$ and $\widetilde{s}_i$, respectively.}
    \label{fig:LSTM_unfold}
\end{figure}

\begin{figure}[ht!]{
    \psfrag{A}[][][0.7]{$Y_i$} \psfrag{B}[][][0.45]{Reference Gen.} \psfrag{C}[][][0.7]{$\widetilde{Y}_i$} \psfrag{D}[][][0.7]{Modified} \psfrag{E}[][][0.65]{$S_i$} \psfrag{F}[][][0.65]{$\widetilde{S}_i$ }\psfrag{G}[][][0.7]{Dense} \psfrag{H}[][][0.8]{Dense} \psfrag{I}[][][0.6]{$ \mathsf{T}_{\theta_{Y}}(\widetilde{Y}_i|Y^{i-1})$ }\psfrag{J}[][][0.6]{$ \mathsf{T}_{\theta_{Y}}(Y_i|Y^{i-1})$} \psfrag{K}[][][0.85]{DV} \psfrag{L}[][][0.7]{LSTM} \psfrag{M}[][][0.7]{Layer} \psfrag{N}[][][0.7]{Input} \psfrag{Z}[][][0.7]{$\widehat{D}_{Y}(\theta_{Y}, \mathcal{D}_n)$}
    \centerline{\includegraphics[scale=0.35]{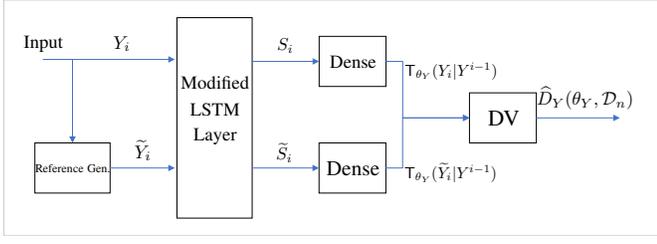}}}
    \caption{End-to-end architecture for estimating $\widehat{D}_{Y}(\theta_{Y}, \mathcal{D}_n)$. For each batch of time sequences, a batch of the same size is sampled from the reference measure. Together, these samples are fed into the NN to compute $\mathsf{T}_{\theta_Y}$ and $\mathsf{T}_{\theta_{Y \| X}}$, from which the estimate is assumbled.}
    \label{fig:DINE}
\end{figure}

We modify the structure of the LSTM to perform the calculations:
\begin{equation}
\begin{aligned}
    s_i = F(y_i, s_{i-1}) = s(y_i|y^{i-1})\\
    \widetilde{s}_i = F(\widetilde{y}_i, s_{i-1}) = s(\widetilde{y_i}|y^{i-1})\label{EQ:modified_LSTM}
\end{aligned}
\end{equation}
A similar modification is introduced for $\widehat{D}_{Y \| X}$ by substitution of $y_i$ with $(y_i,x_i)$ and $\widetilde{y}_i$ with $(\widetilde{y}_i,x_i)$, we have:
\begin{equation}
\begin{aligned}
    s_i = F(y_i,x_i, s_{i-1}) = s(y_i|y^{i-1},x^i)\\
    \widetilde{s}_i = F(\widetilde{y}_i,x_i s_{i-1}) = s(\widetilde{y}_i|y^{i-1},x^i).
\end{aligned}
\end{equation}
A visualization of a modified LSTM cell (unrolled) is shown in Fig. \ref{fig:LSTM_unfold}. The LSTM cell's output is the sequence  $\{ (s_i, \widetilde{s}_i) \}_{i=1}^n$, which is fed into a fully-connected layer to obtain $\mathsf{T}_{\theta_Y}$ and $\mathsf{T}_{\theta_{Y \| X}}$.
As demonstrated by Algorithm \ref{DINE_algorithm} and Fig. \ref{fig:DINE}, in each iteration we draw $\mathcal{D}_B$, a subset on $\mathcal{D}_n$, of size $B$. We feed the NN with $\mathcal{D}_B$ to acquire $ \mathsf{T}_{\theta_{Y}}$, $ \mathsf{T}_{\theta_{Y\| X}}$. Those enter the NN loss function \eqref{EQ:estimator_explicit}, and gradients are calculated to update the NN parameters $\theta_Y, \theta_{Y \| X}$.

\subsection{Neural Distribution Transformer}
\begin{figure}[ht]{
    \psfrag{A}[][][0.85]{LSTM} \psfrag{B}[][][0.8]{Dense} \psfrag{C}[][][0.8]{Dense} \psfrag{D}[][][0.67]{Power} \psfrag{E}[][][0.7]{$X_i$} \psfrag{F}[][][0.7]{$Y_{i-1}$}\psfrag{G}[][][0.7]{$N_i$} \psfrag{H}[][][0.55]{$\nabla\widehat{I}_{\mathcal{D}_n}(X^n \rightarrow Y^n)$} 
    \psfrag{I}[][][0.67]{Constraint} 
    \centerline{\includegraphics[scale=0.42]{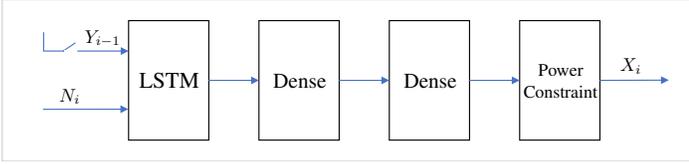}}}
    \caption{The NDT. The noise and past channel output (if feedback is applied) are fed into an NN. The last layer performs normalization to obey the power constraint, if needed.}
    \label{fig:NDT}
\end{figure}

The DINE model is an effective approach to estimate the argument of \eqref{EQ:CFB}. 
However, finding the capacity comprises maximization of the DI with respect to the input distribution. 
For this purpose we present the NDT model that represents a general input distribution of the channel.
At each iteration $i=1,\dots ,n$ the NDT maps an i.i.d noise vector $N^i$ to a channel input variable $X_i$.
When feedback is present the NDT maps $(N^i,Y^{i-1}) \longmapsto X_i$.
Thus, NDT is represented by an RNN with parameters $\mu$ as shown in Fig. \ref{fig:NDT}.
The NDT model is used to generate the channel input $X^n$, and the DINE estimates the DI between $X^n$ and $Y^n$.

\subsection{Complete Architecture Layout}
Combining DINE and NDT models into a complete system enables capacity estimation. As shown in Fig. \ref{fig:complete_system}, the NDT model is fed with i.i.d. noise and its output is the samples $X^n$. These samples are fed into the channel to generate outputs. Then, DINE uses $(X^n,Y^n)$ to produce the estimate $\widehat{I}_{\mathcal{D}_n}(\mathcal{X}\rightarrow \mathcal{Y})$. To estimate capacity, DINE and NDT models are trained together. The training scheme, as shown in Algorithm \ref{alg:cap_est}, is a variant of alternated maximization procedure. This procedure iterates between updating the DINE parameters $\theta$ and the NDT parameters $\mu$, each time keeping one of the models fixed. At the end of training a long Monte-Carlo evaluation of $\sim10^6$ samples is done in order to estimate the expectations in \eqref{EQ:estimator_explicit}.

\begin{algorithm}[ht]
\caption{Capacity Estimation}
\label{alg:cap_est}
\textbf{input:} Continuous channel, feedback indicator\\
\textbf{output:} $\widehat{I}_{\mathcal{D}_n}(\mathcal{X}\rightarrow \mathcal{Y}, \mu)$, estimated capacity.
\algrule
\begin{algorithmic}
\State Initialize DINE parameters, $\theta_Y, \theta_{Y \| X}$
\State Initialize NDT parameters $\mu$
    \If{feedback indicator}
        \State Add feedback to NDT
    \EndIf
\Repeat
\State \textbf{Step 1: Train DINE model}
\State Generate B sequences of length T of i.i.d random noise 
\State Compute $\mathcal{D}_B = \{ (x_{i}^{T},y_{i}^{T}) \}_{i=1}^B$ with NDT and channel
\State Compute $\widehat{D}_{Y \| X}(\theta_{Y\|X}, \mathcal{D}_B)$, $\widehat{D}_{Y}(\theta_{Y}, \mathcal{D}_B)$
\State Update DINE parameters: \\
\hspace{\algorithmicindent}\hspace{\algorithmicindent}$\theta_{Y\| X} \leftarrow \theta_{Y\| X} + \nabla\widehat{D}_{Y \| X}(\theta_{Y\|X}, \mathcal{D}_B)$ \\
\hspace{\algorithmicindent}\hspace{\algorithmicindent}$\theta_{Y} \leftarrow \theta_{Y} + \nabla\widehat{D}_{Y}(\theta_{Y}, \mathcal{D}_B)$
\State \textbf{Step 2: Train NDT}
\State Generate B sequences of length T of i.i.d random noise 
\State Compute $\mathcal{D}_B = \{ (x_{i}^{T},y_{i}^{T}) \}_{i=1}^B$ with NDT and channel
\State compute the objective: \\
\hspace{\algorithmicindent}$\widehat{I}_{\mathcal{D}_B}(\mathcal{X}\rightarrow \mathcal{Y}, \mu) = \widehat{D}_{Y \| X}(\theta_{Y\|X}, \mathcal{D}_B) - \widehat{D}_{Y}(\theta_{Y}, \mathcal{D}_B)$
\State Update NDT parameters: \\
\hspace{\algorithmicindent}\hspace{\algorithmicindent}$\mu \leftarrow \mu + \nabla_\mu\widehat{I}_{\mathcal{D}_B}(\mathcal{X}\rightarrow \mathcal{Y}, \mu) $
\Until{convergence}
\State Monte Carlo evaluation of  $\widehat{I}_{\mathcal{D}_n}(\mathcal{X}\rightarrow \mathcal{Y}, \mu)$ \\
\Return $\widehat{I}_{\mathcal{D}_n}(\mathcal{X}\rightarrow \mathcal{Y}, \mu)$
\end{algorithmic}
\end{algorithm}

Applying this algorithm to channels with memory estimates their capacity without any specific knowledge of the channel underlying distribution.
Next, we demonstrate the effectiveness of this algorithm on continuous alphabet channels.

\section{Numerical Results}
We demonstrate the performance of Algorithm \ref{alg:cap_est} on the AWGN channel and the first order MA-AGN channel. The numerical results are then compared with the analytic solution to verify the effectiveness of the proposed method.

\subsection{AWGN channel}
\begin{figure}[!ht]{
    \centerline{\includegraphics[width=0.85\linewidth]{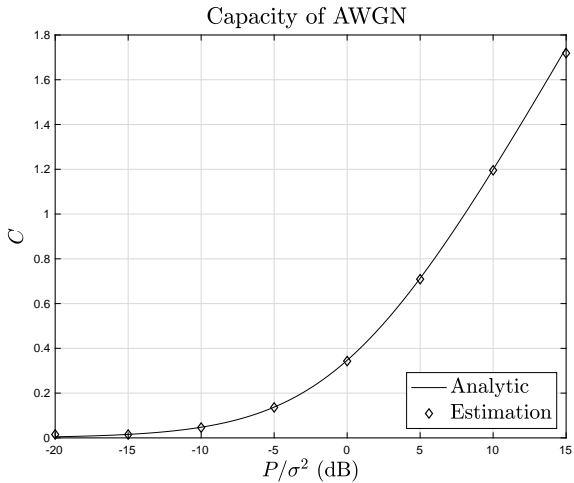}}}
    \caption{Estimation of AWGN channel capacity for various SNR values}
    \label{fig:awgn_results}
\end{figure}

The power constrained AWGN channel is considered. This is an instance of a memoryless, continuous-alphabet channel for which analytic solution is known. The channel model is
\begin{equation}
    Y_i = X_i + Z_i,  \;\; i \in \mathbb{N},
\end{equation}
where $Z_i \sim \mathcal{N}\left(0, \sigma^2\right)$ are i.i.d RVs, and $ X_i $ is the channel input sequence bound to the power constraint $\mathbb
E\left[ X_i^2 \right] \leq P$. The capacity of this channel is given by $\mathsf{C} = \frac{1}{2}\log\left(1+\frac{P}{\sigma^2}\right)$. In our implementation we chose $\sigma^2 = 1$ and estimated capacity for a range of $P$ values. The numerical results are compared to the analytic solution in Fig. \ref{fig:awgn_results}, where a clear correspondence is seen.

\subsection{Gaussian MA(1) channel}

We consider both the FB ($C_{\mathsf{FB}}$) and the FF ($C_{\mathsf{FB}}$) capacity of the MA(1) Gaussian channel. The model here is:
\begin{align}
    &Z_i = \alpha U_{i-1} + U_i \nonumber\\
    &Y_i = X_i + Z_i
    \label{EQ:MA_1_model}
\end{align}
where, $U_i \sim\mathcal{N}(0, 1)$ are i.i.d., $X_i$ is the channel input sequence bound to the power constraint $\mathbb
E\left[ X_i^2 \right] \leq P$, and $Y_i$ is the channel output.

\vspace{1.5mm}
\subsubsection{\underline{Feedforward capacity}} 
\begin{figure}[!ht]{
    \centerline{\includegraphics[width=0.85\linewidth]{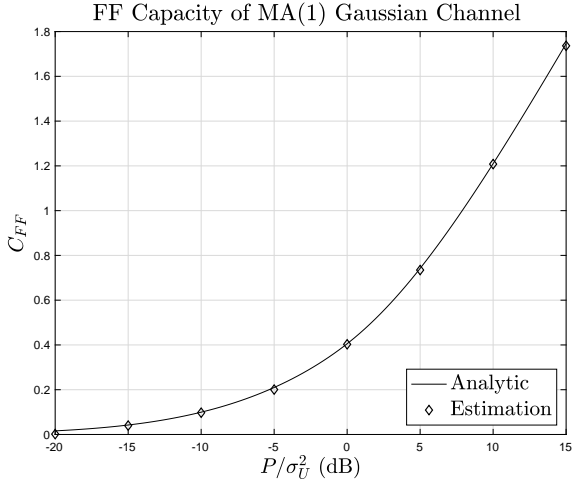}}}
    \caption{Performance of $C_{\mathsf{FF}}$ estimation in the MA(1)-AGN channel.}
    \label{fig:C_ff}
\end{figure}

The FF capacity of the MA(1) Gaussian channel with input power constraint can be obtained via the water-filing algorithm \cite{BOOK:Cover}. This is the benchmark against which we compare the quality of the $C_{\mathsf{FF}}$ estimate produced by Algorithm \ref{alg:cap_est}. Results are shown in Fig. \ref{fig:C_ff}.  

\vspace{1.5mm}

\subsubsection{\underline{Feedback capacity}}
\begin{figure}[!ht]{
    \centerline{\includegraphics[width=0.85\linewidth]{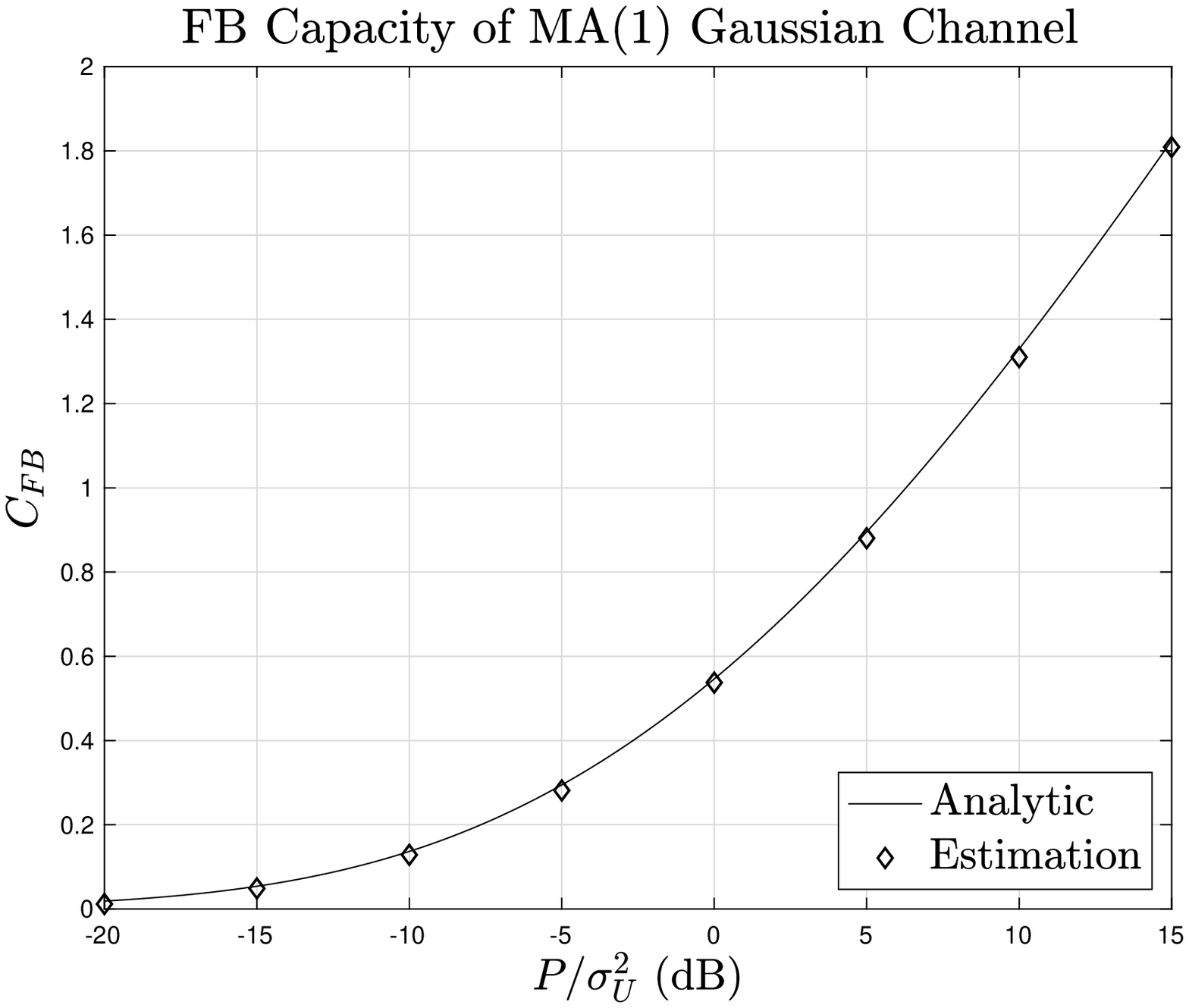}}}
    \caption{Preformance of $C_{\mathsf{FB}}$ estimation in the MA(1)-AGN channel.}
    \label{fig:C_fb}
\end{figure}

\begin{figure}[!ht]{
    \centerline{\includegraphics[width=0.85\linewidth]{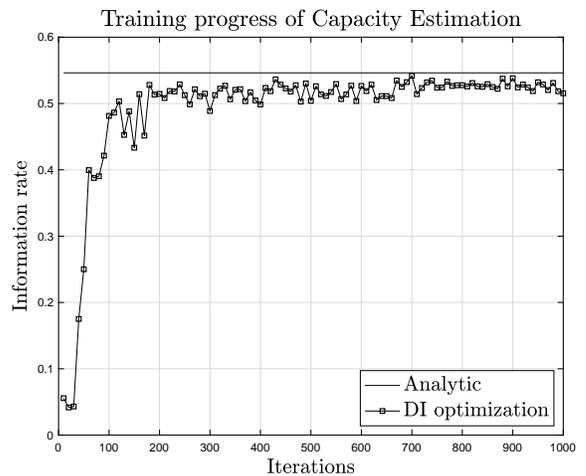}}}
    \caption{Optimization progress of DI rate of Algorithm \ref{alg:cap_est} for the FB setting with $P=1$. The information rates were estimated by a Monte-Carlo evaluation of \eqref{EQ:DI_est} with $10^5$ samples.}
    \label{fig:covergence}
\end{figure}

Computing the FB capacity of the ARMA(k) Gaussian channel can be formulated as a dynamic programming, which is then solved via an iterative algorithm \cite{PAPER:ARMA_Yang}. For the particular case of \eqref{EQ:MA_1_model}, $C_{\mathsf{FB}}$ is given by $-\log(x_0)$, where $x_0$ is a solution to a 4th order polynomial equation. The estimates for $C_{\mathsf{FB}}$ produced by Algorithm \ref{alg:cap_est} are compared to the analytic solutions in Fig. \ref{fig:C_fb}. The optimization dynamics for our algorithm are shown in Fig. \ref{fig:covergence}.

\vspace{1.5mm}
\section{Conclusion and Future Work}
We presented a methodology for estimating FF and FB capacities that uses the channel as a black-box, i.e., without assuming the channel model is known and only relying its output samples. The main building block were a novel DI estimator (DINE) and the NDT model, both implemented based on RNNs. The performance of the estimator was tested on AWGN and MA(1)-AGN channels, showing estimates that agree well with analytic solution. 

Despite the empirical effectiveness of DINE, we stress that it is neither a lower nor a upper bound on the true DI (see \eqref{EQ:DI_entropy}-\eqref{EQ:Entropy_decomposition}). A main goal going forward is to revise DINE so that is provably lower bounds the true value. This will imply that the induced capacity estimator lower bounds the theoretical fundamental limit. Extension of our method to multiuser channels is also of interest, as capacity results in multiuser information theory are quite scarce. Another objective is coupling DINE with theoretical performance guarantees.


\end{document}